# Do tar roads bring tourism? Growth corridor policy and tourism development in the Zambezi region, Namibia

*Linus Kalvelage, Javier Revilla Diez, Michael Bollig*

*Abstract*

*There are high aspirations to foster growth in Namibia's Zambezi region via the development of tourism. The Zambezi region is a core element of the Kavango-Zambezi Transfrontier Conservation Area (KAZA), a mosaic of areas with varying degrees of protection, which is designed to combine nature conservation and rural development. These conservation areas serve as a resource base for wildlife tourism, and growth corridor policy aims to integrate the region into tourism global production networks (GPNs) by means of infrastructure development. Despite the increasing popularity of growth corridors, little is known about the effectiveness of this development strategy at local level. The mixed-methods approach reveals that the improvement of infrastructure has led to increased tourism in the region. However, the establishment of a territorial conservation imaginary that results in the designation of conservation areas is a necessary precondition for such a development. Despite the far-reaching territorial claims associated with tourism, the benefits for rural residents are limited.*

## 1. Introduction

The promotion of tourism is a central pillar of Namibia's economic development strategy. Since the 1990s, the designation of nature conservation areas has been envisioned to protect wildlife and wilderness landscapes while at the same time boosting growth in rural areas. This has led to the emergence of a resource base for wildlife tourism. Growth corridors have been a spatial planning tool for decades, with the aim of fostering economic development in the hinterland through investment in infrastructure (Dannenberg et al., 2018). Recent growth corridor policies incorporate tourism as a development strategy, but it is not clear whether local residents benefit from this approach.

Windhoek is connected by a growth corridor to a tourism destination that is increasingly gaining popularity: the Zambezi region in north-eastern Namibia. The Walvis Bay Ndola Lubumbashi Development Corridor (WBNLDC) is the "new generation growth corridor" that is based upon its predecessor, the Trans-Caprivi Corridor (TCC). While the latter was limited to investments in infrastructure and the smoothing of logistic procedures, the WBNLDC is designed to incorporate more advanced spatial development policies, such as the creation of hubs, gateways and targeted value-chain promotion. Tourism is a proclaimed means of fostering economic growth. In the Zambezi region, this vision meets a partly competing, partly synergetic vision of the future, which is promoted by a network of transnational actors aiming to create one of the world's largest nature conservation landscapes, the Kavango-Zambezi Transfrontier Conservation Area. Both visions bear the promise that conservation policies and infrastructure connectivity will increase gains for local residents. This contribution aims to examine this claim.

In conceptual terms, the analysis is influenced by the current literature on global production networks. While this concept is gaining popularity for exploring uneven development outcomes, there are two points that remain largely overlooked: the role of infrastructure in GPNs and the territorialisation of GPNs in the resource region. In this paper it is argued that corridor policies are designed to foster economic growth by coupling regions to GPNs. The wildlife tourism GPN relies on conservation territories as their resource base. The resources underlying the tourism GPN, wildlife and wilderness landscapes, are place-bound. These resources therefore first require access to infrastructure in order to untap their economic potential and, second, territorialisation strategies in the resource region are needed to maintain the resource base for the GPN. Whether these developments are beneficial for the region depends on the degree of value that can be captured locally according to GPN researchers (Henderson et al., 2002).

The questions addressed in this paper are therefore whether growth corridors succeed in enhancing value creation from tourism in the hinterland and, more importantly, whether the created value can be appropriated by the local residents.

Applying a historical perspective, first the parallel evolution of conservation policy and infrastructure development is shown. Second, traffic census data are analysed to assess the extent to which infrastructure improvements have had the desired effect of fostering tourism-related traffic to the region. Third, it is investigated whether positive development effects reach local residents. This is done by presenting data from a recent household survey. This will lead to a refined understanding of the territorialisation of tourism GPNs and the effectiveness of growth corridor policy in fostering tourism-driven development.

## 2. Theoretical background
### a. Tourism GPNs for regional development

In many countries of the world, wildlife tourism is regarded by conservationists and development planners as the ideal solution for sustainable development, stimulating the poor rural population and ensuring the preservation of ecosystems. Advocates of such an approach claim that the tourism industry is a rapidly growing sector, especially in developing countries, with the potential to diversify the economy in poor rural areas beyond agricultural production (Scheyvens, 2007). The tourism sector comprises different industries, for example accommodation, food and beverages, transportation, culture, sports and recreational services, thus ensuring a wide spread of indirect effects (Newfarmer et al., 2018). However, critics warn that in many destinations the formation of tourism enclaves leads to the exclusion of local residents from the benefits (Mbaiwa, 2017), revenues are lost due to local leakages (Sandbrook, 2010) and tourism can have negative impacts on the ecological system (Stronza et al., 2019). There is also criticism that globalised tourism is vulnerable to external shocks, as the current pandemic has impressively shown (Lendelvo et al., 2020).

An increasingly popular tool used to tackle these uncertainties is the application of a GPN approach to tourism. Global production networks, a conceptual expansion of the GVC approach applying a network heuristic, is able to grasp the complexity of the industry and aims to analyse notions of power, embeddedness and value in globalised modes of production (Henderson et al., 2002). In an insightful study on Bali and Sulawesi, Nanda and Hargreaves show how external shocks can lead to a restructuring of the tourism GPN (2013). The role of gender and race in upgrading dynamics in the Kenyan safari tourism sector has been examined (Christian, 2016). Murphy explores the tourism GPN in Zanzibar and finds that local enterprises are increasingly marginalised while foreign enterprises capture large shares of the value (2019).

In the safari tourism GPN, global tour operators in outbound countries act as lead firms due to their capacities for bundling services, their direct access to the customer and their ability to govern the network (Christian, 2016). National tour operators act as destination management operators, bundling services like accommodation, domestic transport and excursions and selling these packages to the lead firm. Daly and Gereffi (2017) analyse different distribution channels in Africa, distinguishing between direct booking (the consumer books with a service provider), online packages (the consumer books via an online portal that uses global distribution systems to place bookings with service providers) and package booking (the consumer books with a travel agent, who purchases packages from GTOs that are bundled by inbound tour-operating companies). However, this specific network configuration differs considerably from a second form of tourism that plays a major role in Southern Africa: hunting tourism. In Namibia, hunting tourism is dominated by domestic actors that bundle, operate and sell the tour packages directly to customers in the outbound countries (Kalvelage et al., 2020). Windhoek, as the gateway city, is crucial for enhancing the value of the nature in the hinterland, although government policies ensure that the resource region participates in the value distribution (ibid.).

### b. Growth corridors and tourism development

Jaffee (2019) has argued that city-regions strategically use large-scale infrastructure investments to exploit and expand geographic and physical assets and in turn to capture economic benefits from GVCs. In Southern Africa, a multitude of development corridors have emerged during the past two decades in parallel to the ongoing economic integration of the Southern African Development Community. Backed by international organisations such as the IMF and the World Bank, countries have increasingly adopted the spatial development instrument (SDI) of growth or development corridors to foster economic growth (Dannenberg et al., 2018). The idea is to combine infrastructure development with targeted interventions to promote specific sectors (Nogales, 2014). The formation of multi-stakeholder alliances aims to create a critical mass of investment in order to boost the economy in specific locations by co-location effects (ibid.). Beside other sectors like mining, agriculture and manufacturing, tourism is one of the industries that is expected to exhibit growth potential. The Maputo Corridor in South Africa, for instance, integrated tourism into its planning early on (Rogerson, 2001). Transport infrastructure is a strong determinant of a destination's attractiveness and thus also of tourism-led development (Khadaroo and Seetanah, 2007).

Development corridors are both tangible and intangible: a network of roads, railways, pipelines and ports is accompanied by regulatory reforms with the aim of ensuring the free circulation of commodities, capital and people between production sites and economic hubs (Enns, 2018). While previous development corridors were based on envisaged neoclassical infrastructure effects for development, more recent approaches are oriented towards the GVC literature and aim to create a favourable environment for economic activities alongside the infrastructure development projects (Dannenberg et al., 2018). Thus, development corridors serve to connect resources in the hinterland of economic hubs to global production networks (Sen, 2014). As Hesse (2020) states, logistics are "a vital component of the making of territories in a networked economy", as they are crucial for coordinating the flow of commodities, and a connection to or a disconnection from logistics can lead to variegated development outcomes. Spatial development initiatives come with a territorial claim: by expressing "desirable futures" of modernity, alternative uses of space are displaced (Müller-Mahn, 2019). This may have very tangible outcomes, as local residents can be displaced to make space for corridor development (Enns, 2018). Current spatial development initiatives to install growth corridors or development corridors in resource-rich countries are strategies to gain access to and create value from resources in the hinterland.

### c. Territoriality and territorialisation in nature-based GPNs

The conceptual advances of GPNs with respect to the role of non-firm actors in the formation and governance of global production processes are widely acknowledged in the literature (e.g. Bridge 2008, Henderson et al., 2002). However, territorialisation processes associated with the coupling of regions into GPNs remain largely unexplored. While the GPN approach originated in research on the manufacturing industry, the concept has been expanded in recent years to cover sectors linked more directly to natural resources, including extractive industries (Breul et al., 2018), fish production (Irarrázaval and Bustos-Gallardo, 2018), wood (Gibson and Warren, 2016) and wildlife tourism (Kalvelage et al., 2020). These GPNs have a specific notion of territoriality and materiality that affects the network configuration and, ultimately, development outcomes (Bridge and Bradshaw, 2017): resource-driven GPNs are less flexible in spatial terms, as they depend on processes of production prior to human labour (Bridge, 2008). Therefore, resource-driven GPNs are "embedded within state structures to a much greater degree" (Bridge, 2008:413), and this territoriality has implications for the distribution of value across places. The materiality of resources implies that value creation is driven by ownership of resources and therefore that there is an incentive to control the space where these resources are located.

The discovery or invention of new resources is accompanied by frontier dynamics that "dissolve existing social orders" (Rasmussen and Lund, 2018:388). Territorialisation sets in as a strategy to gain control over resources in these resource frontiers. Territorialisation is "the ambition to control a geographic area, involving a series of operations: establishing a territorial administration, instituting a legal system, establishing boundaries and ensuring the capacity to enforce all of this" (Rasmussen and Lund, 2018:393). In the context of wildlife tourism, territorialisation is crucial for resource production. This holds true at enterprise level, where tourism resorts form enclave structures that prevent outsiders from accessing it (Saarinen, 2017). In a similar vein, research has shown the role of cities in the Global South in territorialising their hinterlands to access newly emerging resources, such as ecosystem service certificates (Schindler and Kanai, 2019). Territorialisation is also visible in the formation of nature conservation areas, where ecotourism companies "employ different techniques of government to secure business-friendly environments and territories" (Bluwstein, 2017:101). One of these techniques is the creation of a regional territorial imaginary that can encourage resident, corporate and state actors to support a space committed to green development (Mendoza et al., 2017). This multi-scalar, networked mode of environmental governance has been analysed using the GPN approach in the case of ecosystem services in the Amazon (Urzedo et al., 2020). In Namibia's Zambezi region, zonation plans demarcate areas that are used exclusively for tourist activities or that are reserved for wildlife (Kalvelage et al., 2020).

However, without infrastructure to facilitate access there is no resource. While territorialisation is a precondition by securing resource access to tourism GPNs, infrastructure is needed for the commodification of wildlife. In contrast to other industries, in tourism the consumption occurs at the production site. In order to circulate the tourism product as a commodity, access to infrastructure has to be developed. This paper will show how the development of infrastructure access and a conservation narrative paced the way for the exploitation of wildlife and landscapes as a resource for the tourism industry. As a result of the region's coupling, the tourism GPN makes far-reaching territorial claims.

3. **Methods**

The data for this paper are based on a mixed-methods approach that includes a review of secondary sources, such as websites, policy reports and existing scholarly literature, a business survey conducted among Zambezi tourism enterprises, qualitative interviews with key stakeholders in the Namibian tourism industry, a traffic census carried out in July 2019 and a household survey from 2019. The business survey was used to collect data at enterprise level and the qualitative interviews with key actors in the tourism GPN conducted in parallel provided background information useful for interpreting the results. These findings were supplemented with household-level data to gain an understanding of the impact of tourism on residents. An innovative approach was needed to measure the effects of the infrastructure development, which led to the application of a traffic census. The most important reason for this multi-perspective approach is that tourism-related data at regional level in Namibia are scarce. Therefore, the triangulation of data made it possible to portray the complexity of the situation. Second, combining survey data with qualitative data is a good way to explore whether the findings can be projected to a region to establish a regional pattern (cf. *ethnographic upscaling*, Bollig, Schnegg & Menestrey-Schwieger 2020).

The two main publications on the history of the Zambezi region (Kangumu, 2011; Lenggenhager, 2015) served as a point of departure to trace back the development of the corridor and the tourism sector in the region. A review of scientific and government reports from the 1980s and 1990s facilitated a reconstruction of the development of the region's tourism sector. Details on the corridor plans were added by analysing policy plans and reports.

A traffic census was used as a tool to measure the impact of the road on tourism. The corridor enters the region on the western edge and leads north to Zambia. There is another gateway to the Zambezi region in Ngoma, where the road crosses into Botswana. A team of 9 enumerators collected traffic data on three days in July/ August 2019 (July 29, July 31, August 2), four at the Wenela border post (2 for each direction), four at the Ngoma border post and one in Kongola (cf. figure 3). Between 6 a.m. and 6 p.m., these teams counted all vehicles entering and leaving the region, collecting a variety of data on each vehicle: the origin of the number plate, number of passengers, branding on the car, the type of car and the cargo transported by trucks. The data were collected using Survey Solutions, a free-of-charge survey tool provided by the World Bank. The data collection form was installed on tablets on which the data were stored temporarily until they could be uploaded to the server whenever there was access to the network. This approach had several shortcomings: first, July and August are the peak of the tourist season in the region, so the counts are not representative of the whole year. Second, the census might include double counts, for instance cars that passed two data collection points. Third, on some occasions when the traffic was dense, the enumerators were unable to collect all the information in detail due to time limitations. However, overall the approach proved successful. The findings can be used to answer a variety of questions related to the economic effect of growth corridors.

The household survey was conducted by the collaborative research centre "Future Rural Africa" (www.crc228.de). In Namibia, 652 households were surveyed, comprising 3271 household members. The sampling covered the entire Zambezi region without the urban centre (Katima Mulilo). The sampling strategy was a two-stage, stratified random sampling. First, all the rural enumeration areas were classified using three land-use categories: mainly conservation, mainly intensification and other. Out of a total of 292 enumeration areas, 45 were sampled randomly, from which 15 households were then randomly selected for surveying. The household representatives were interviewed using a questionnaire that covered a wide range of topics, including a section on the household's income, assets and expenditure. The interviews were conducted with the help of local assistants, who were able to translate the English questionnaire into the respondents' mother tongue.

The business survey gathered general enterprise data as well as information on employment figures, booking procedures, supply chains and expenditure. 33 of the 47 firms completed the factsheet. Finally, in order to detect causal explanations for the survey and census data, qualitative interviews were conducted with key stakeholders of the Namibian tourism industry during two fieldwork phases from August to November 2018 and from June to August 2019. The stakeholders included lodge operators, professional hunters and conservancy managers in the Zambezi region as well as tour operators and government officials in Windhoek (a total of 65 interviews). While all the information gathered served as background information for interpreting the data, only few of the interviews are directly referenced using the following codes: tour operator (TO), lodge manager (LOD), business association representative (BA), professional hunter (PH).

## 4. Results

### *a. Accessing a resource frontier: corridor development in the Zambezi region*

Located in the north eastern periphery of Namibia, the Zambezi region has been regarded as a resource frontier ever since the arrival of European settlers to Southern Africa. Formerly known as the Caprivi strip, the motive for adding the region to the colonial acquisitions in South-West Africa was its presumed value as a transport corridor to the eastern parts of the continent. The dispossession of land and the establishment of white settler farms in Central Namibia was a rapid process starting in 1884. In 1907, the German colonial administration proclaimed that policing should be restricted to the "sphere of influence of the railway line or main roads" (Werner, 1993: 193) which did not include the Caprivi. Between 1890 and 1909, the Eastern Caprivi strip functioned as "an El Dorado for shady characters, criminals or prisoners who went into hiding and a happy hunting ground for both part time and professional trophy hunters" (Kangumu, 2011:132). Game was abundant, as hunting was previously controlled by Paramount Chief Lewanika, who lived in Western Zambia and used the area as a private hunting reserve (Kangumu, 2011). Although Grootfontein was connected to the railway system in 1908, the Caprivi was still difficult to access, with the result that "the German Resident" in Katima Mulilo lived "as in exile" (Meyer, 1910:279). It was only in 1909 that Kurt Streitwolf, a German captain, was installed as *Kaiserlicher Resident* in Schuckmannsburg in order to extend German colonial administration to the Caprivi (Curson, 1947). However, this administration only lasted until 1914, when the Caprivi was seized by Southern Rhodesian troops and administered by the High Commissioner of the Bechuanaland Protectorate (ibid.).

Although the administration was formally handed over to the South-West Africa Protectorate authorities in Windhoek in 1930, the inaccessibility of the region made it necessary for administrative duties to be handled by the *Native Affairs Department* in Pretoria from 1939 onwards. By then the Caprivi could be accessed by train, bus or boat from Livingstone or Kasane, or by plane (Curson, 1945). As early as 1945 the development of a tourist industry was identified as a potential for growth in the region, besides the exploitation of timber, commercial crop farming and logistics on the Zambezi River (ibid.).

Under South African rule, the Odendaal Commission recommended government-driven development, which resulted in an upgrading of infrastructure, including the development of unpaved road connections to Western Caprivi and Ngoma (Zeller, 2009). The region gained military importance due to ongoing clashes with liberation forces in Angola and Zambia during the 1960s and 1970s (Lenggenhager, 2015), which led to further investment in infrastructure, for example the construction of the Mpacha military airport near Katima Mulilo in 1965. Parallel to the infrastructure development, conservation areas were declared: Western Caprivi was proclaimed a Nature Park in 1963 and in 1964 Katima Mulilo and its surroundings were granted the status of a nature reserve (Kangumu, 2011). While first resettlements for the creation of conservation areas date back to the 1930s, the establishment of a state forest and the development of game reserves, Nkasa Rupara and Mudumu, caused further relocations during the 1970s and 1980s (Bollig and Vehrs, 2020). The latter two areas were designated as nature reserves in 1989, setting a milestone for the creation of "an anthropogenic wilderness (ibid.: 34)" that serves the vision of an economically productive conservation landscape. However, the economic potential of the wildlife was not fully exploited until the region was connected to the rest of the country.

After independence, the construction of the TCC was planned to overcome regional disparities caused by the colonial system. The Caprivi retained a peripheral status until the road connecting it with the rest of Namibia was tarred from the mid-1990s and officially opened in 1999 (Zeller, 2009). The construction of a bridge spanning the Zambezi River and connecting Namibia with Zambia marked the termination of the Trans-Caprivi Corridor in 2004.

Table 1 shows the increase in the number of tourism establishments in the Caprivi. Prior to independence in 1990, the number of lodging facilities was distinctly low. During the 1980s, the centre of military conflict shifted westwards, away from the Caprivi, which permitted the emergence of the first camps and fishing lodges. The presence of military forces had caused a depletion of the game population, as officials had hunted excessively, both for sport and to trade ivory (Lenggenhager, 2015). Formalised trophy hunting came into being in 1988, when two concessions enabled PHs from Central Namibia to expand their business to the Caprivi. Yet revenues remained limited and were estimated at 163,000 USD in 1994 (Barnes, 1995).

| Year | 1990* | 1994** | 2005* | 2018*** |
|---|---|---|---|---|
| Accommodation establishments in the Zambezi region | 4 | 8 | 24 | 47 |

*Table 1: Number of accommodation establishments in the Zambezi region.* *Suich & Busch, 2005; **Barnes, 1995; ***Kalvelage et al., 2020.

A change was driven by the introduction of a new conservation institution. In response to more exclusionary conservation policies, CBNRM projects started to emerge across Southern Africa during the 1980s. In Namibia, CBNRM after independence was linked to the political agenda that aimed to overcome territorial disparities caused by colonial administration (Dressler et al., 2010). CBNRM policy permits communities to form a conservancy and to implement conservation measures and grants them the right to market wildlife as a resource for the tourism industry (Kalvelage et al., 2020). The first conservancy to be established in Zambezi was Salambala in 1998, 14 more have emerged since then. Despite the political unrest triggered by the independence movement, known as the "Caprivi conflict" between 1994 and 1999, the number of tourism establishments in Zambezi increased considerably (Table 1).

In 2000, the Walvis Bay Corridor Group was established to manage four growth corridors connecting the port in Walvis Bay to the landlocked hinterland, including the WBNLDC. The members of the group are stakeholders from Walvis Bay, e.g. Walvis Bay Port Users' Association (WBPUA), logistics companies (Namibia Logistics Association) and ministries. In addition to the development of "hard" infrastructure, like roads, rails, ports, electricity grid, water and ICT, the corridor plan drafted by the Australian consultancy AURECON foresees the instalment of complementary programmes such as truck stops, green-schemes, agri-hubs and logistics parks (cf. figure 1, AURECON, 2014). Furthermore, catalytic investments in key sectors (mining, energy, manufacturing, water, aquaculture, agriculture, property and tourism) are planned with the aim of inducing broader economic stimuli in selected hubs along the corridor. Due to its strategic location on the borders of Namibia, Zambia, Zimbabwe and Botswana, Katima Mulilo is highlighted in the national logistics strategy as possessing "the most viable and unique nodal development opportunities" (Walvis Bay Corridor Group, 2018). Given its vicinity to nature parks and attractions, substantial growth potential is expected for the tourism sector (Ministry of Lands and Resettlement, 2015). Subsequently, the Tourism Investment Strategy encouraged the formation of a public-private partnership for tourism-related waterfront development in Katima Mulilo (Ministry of Environment and Tourism, 2016), which, however, failed to materialise due to maladministration and financial irregularities (https://www.namibian.com.na/148511/archive-read/Zambezi-waterfront-closes-doors).

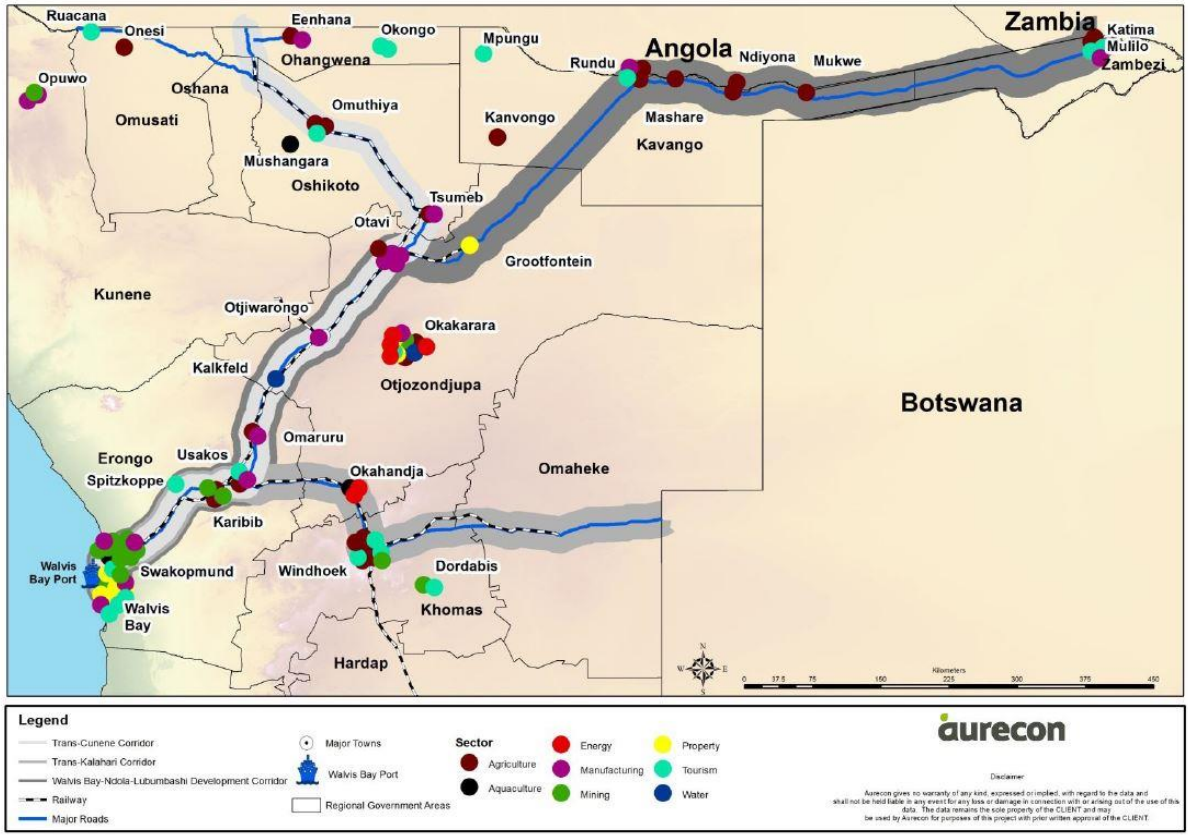

Figure 1: The WBNLDC corridor vision. Source: AURECON 2014

The attention of international donors shifted increasingly towards the concept of trans-frontier conservation areas (TFCA) during the late 1990s and early 2000s (Büscher, 2010). Thus, an international alliance of donors and conservationists pushed ahead the establishment of the Kavango-Zambezi Tranfrontier Conservation area (KAZA), which integrates the Zambezi region into a wider network of nature conservation attempts in the neighbouring countries (cf. figure 4). After an initial memorandum of understanding in 2006, KAZA was finally launched in 2012. The stakeholders include international donors (e.g. German Development Bank, Swiss Agency for Development and Cooperation, World Bank), large conservation organisations (e.g. Peace Parks Foundation, World Wide Fund for Nature, African Wildlife Foundation) and government bodies (Dutch Ministry of Foreign Affairs, Southern African Development Community, ministries of the participating countries). The proclaimed aim of the KAZA initiative is "to sustainably manage the Kavango-Zambezi ecosystem (www.kavangozambezi.org)" and to transform the KAZA region "into a premier tourist destination in Africa" (https://tfcaportal.org/system/files/resources/KAZA%20TFCA%20Treaty_SIGNED.pdf). To this end, administrative units are formed with the aim of working towards a harmonised legal framework.

Parallel to the infrastructure development that connects the Zambezi region to the mainland, a conservation-based territorial imaginary has emerged that has resulted in the formation of a conservation landscape. Today, 54 % of the region's territory is protected to varying degrees, including national parks, communal conservancies, a state forest, tourism priority areas and wildlife corridors (own calculation, cf. figure 2). Both processes have led to nature being classified as a resource, which is linked with the tourism GPN, thus enabling a transfer of value to national gateways and global lead firms (Kalvelage et al., forthcoming). Continuing a strong critique of earlier conservation approaches, these policies are not undisputed: research has revealed discontent among smallholder farmers regarding harvest losses caused by wildlife, residents claim the distribution of conservancy income does not reach individual households and that the designation of areas for tourist activities negatively affects agriculture (Hulke et al., 2020). This calls into question the effectiveness of promoting nature-based tourism as a tool to foster development in rural areas. The remaining two sections of this paper therefore aim to clarify first, whether growth corridors bring tourism, and second, whether the value created via tourism reaches rural households in the region.

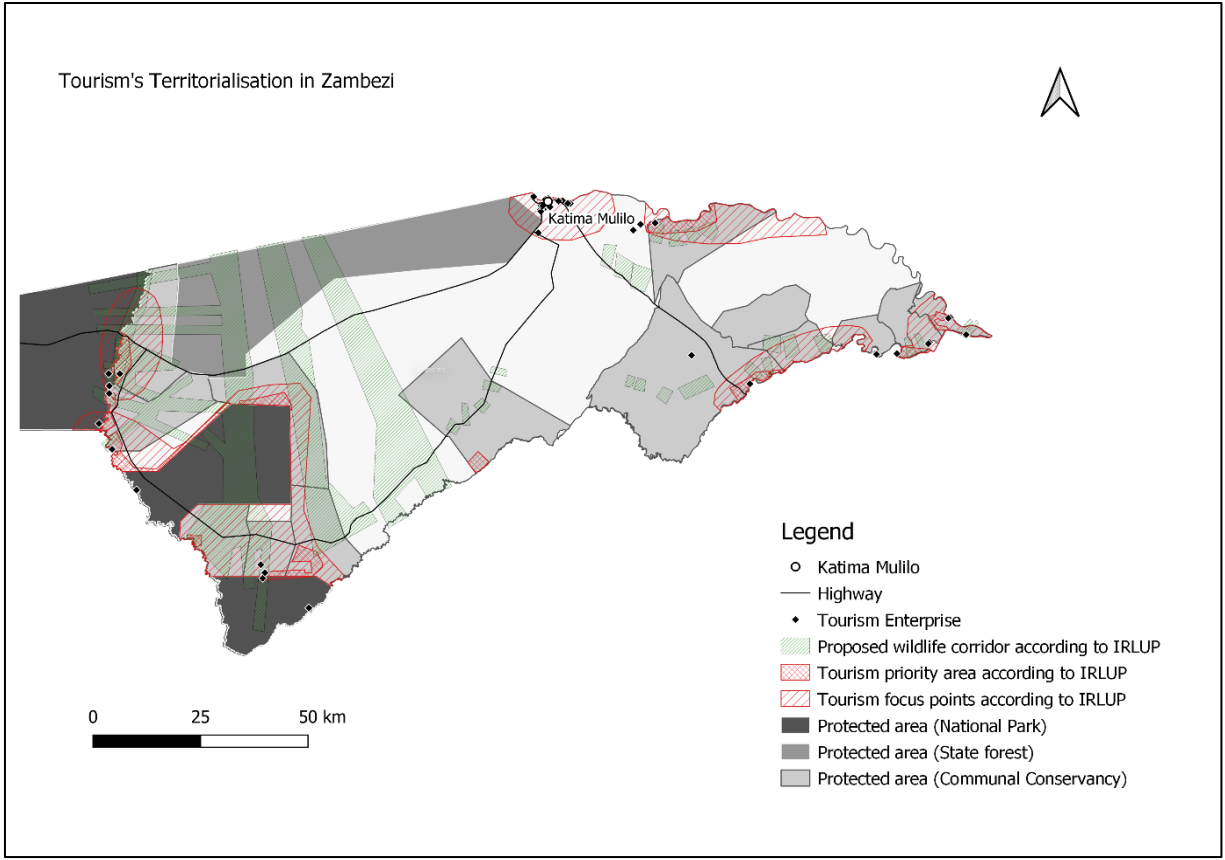

*Figure 2: Territorialisation of the tourism sector in the Zambezi region*

### b. Do tar roads bring tourism? The territoriality of a growth corridor

Based on traffic census data, traffic was classified into three categories: tourism, cargo and other. The first category includes all vehicles with the markings of a tourist car rental, tour buses and self-drive tourists. The second category includes all trucks and cars with a company sign or logo. The remaining vehicles were classified as "other". Figure 3 shows the traffic flows on three days at the different posts. A total of 1795 vehicles were recorded, with tourism-related traffic accounting for 25 %, cargo and other business for 36 % and other traffic for 39 %.

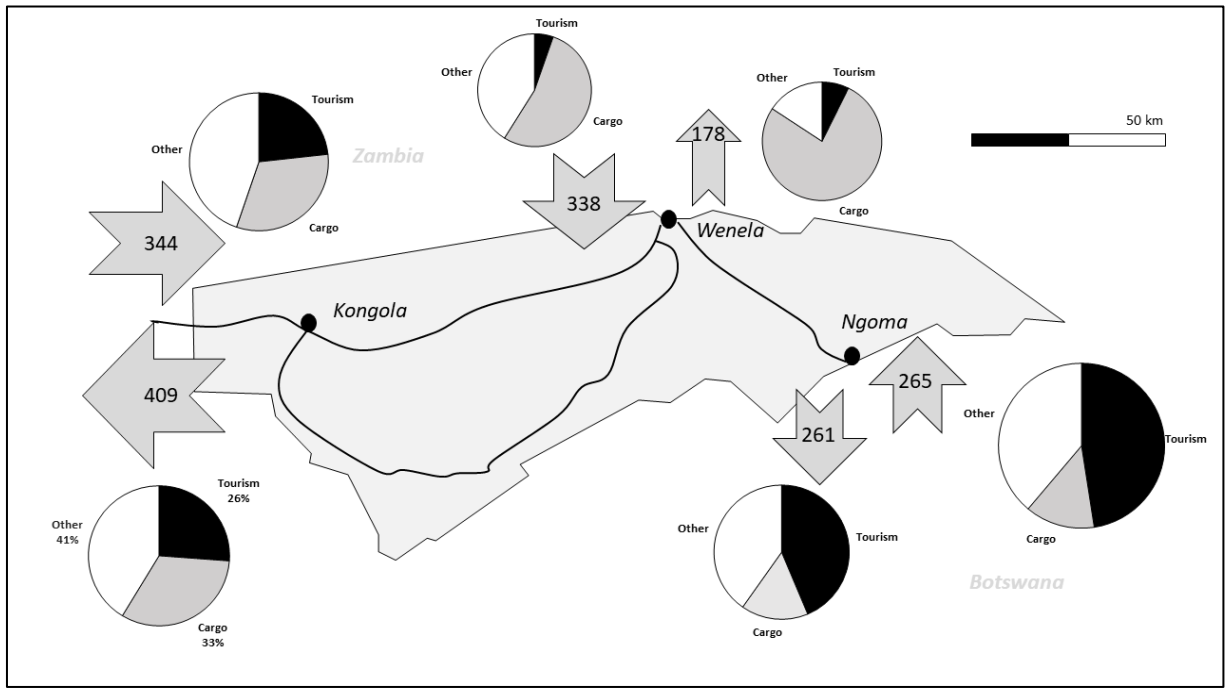

*Figure 3: Results of the traffic census, number in arrows shows total N.*

Figure 3, however, shows the unbalanced distribution of tourist-related traffic. While Wenela did not exhibit a significant number of arrivals and departures, the share of tourist-related traffic was considerably higher in Ngoma. This is not a surprise as the classic tourism route in Zambezi leads to Chobe National Park in Botswana and further onwards to Victoria Falls in Zimbabwe. At Wenela, the border post close to Katima Mulilo, 76 trucks travelled south with 7 containers, 33 loaded with copper, 14 with timber products and 22 with other cargos. Freight traffic to and from the port in Walvis Bay to the mines in Zambia and Congo is significant (LOD1). The poor condition of the roads in Zambia prevents tour operators from offering tours to Zambia (TO1). The quality of the infrastructure is a crucial cost factor for tour operators: *"I think right now [tour operator] is replacing the shock absorbers on every sprinter [Mercedes Sprinter] after every trip, it costs N$ 7,000 every time, so there, infrastructure is very, very important (TO1)"*. Not only tour operators, but also hunting outfitters highlight the importance of infrastructure quality for business. Although most hunting clients arrive at the airport in Katima Mulilo, the road has a crucial function for the acquisition of spare parts and food supplies from Windhoek (PH1).

Figure 4 presents an analysis of the origin of the vehicles recorded in the Zambezi region. 59 % of the vehicles were registered in Namibia, followed by Zambia (20 %), RSA (11 %), Botswana (7 %) and other origins (4 %). The total number of vehicles registered in the Khomas region (479) surpassed even Zambezi (187) by far, followed by the neighbouring Erongo region (142), where the port city of Walvis Bay and the tourist resort of Swakopmund are located. It is striking that the vast majority of the number plates were from Windhoek and the neighbouring Erongo region, the country's economic powerhouse. While this analysis may serve as a rough indicator of the territoriality of the corridor, company headquarters identified by the markings on vehicles can give some indication of how far the impact of the corridor development reaches. The majority of the companies identified are located in the towns along the corridor: Windhoek accounts for 88 companies, followed by Ndola (59), Lusaka (57), Swakopmund (28) and Walvis Bay (22). Most of this traffic consists of trucks transporting primary goods from the copper mines in Northern Zambia and Southern DRC to the coast in Namibia. Windhoek acts as a gateway city for tourism, however, is as well home to a large number of tourism companies: *"(…) the main tourism actually starts here in Windhoek. The people fly mainly all to Windhoek or to*

*Walvis Bay and from here the whole tours start* (BA1)". 22 vehicles from tourism companies based in Katima Mulilo were recorded, as well as vehicles from companies in Kasane (9) and Livingstone (7).

Prior to the tarring of the road, logistics were challenging: *"we started by doing logistics around the Caprivi. Because [...] roads were very bad. I mean a lot of them have been tarred since we opened, and people were scared of coming to north with self-drives because the cell-phone signal was bad [...] (TO2)"*. The improvement of the infrastructure has increased the potential of tourism in the region: "*Well, I guess, since the Trans-Caprivi tar road was finished [tourism has improved], from Rundu to Katima was gravel, a nightmare, 500 km gravel, straight ahead* (TO1)". The good condition of roads in Namibia has been used to market Namibia as a self-drive destination (BA1) and the share of independent travellers has been growing steadily: "*Twenty years ago, nobody dared to come here except by bus. Because they said, you know, I get on a bus, I have a driver who takes me everywhere safely. But at some point they realized, you know, Namibia is so easy to travel. The roads are actually good, the tarred roads* (TO3)". This development has led to an overall increase in traffic, as "b*efore there was a bus with 30 people, now there are 15 cars with two people* (BA1)". However, the Zambezi region remains simply a stopover on the way to the main tourist attractions of Chobe National Park in Botswana and Victoria Falls in Zimbabwe: "*Up in the Zambezi, you can now drive from Rundu to Katima everything on tar and even the loop down there in the corner is already tarred. Did it bring more tourism? It is of course faster tourism* (BA1)". Stakeholders in the region aim to overcome this shortcoming: "(...) *we are working hard with all the accommodation and other bodies here to keep people in the Caprivi for long (...). So our focus has shifted from just that to trying to get more activities and accommodation streamlined to get people to stay here for four days or a week* (TO2)".

A study found that in 2005, 24 establishments catered for an estimated 31,000 guests in the Zambezi region (Suich and Busch, 2005). By 2018, the number of businesses had risen to 47 (Kalvelage et al. 2020), although the total number of guests per year is not clear. In an analysis of border post data, the Ministry of Environment and Tourism (MET) counted 1,499,442 arrivals to Namibia in 2017, with 580,519 arrivals reporting that their trip was for holiday purposes (cf. table 2). This figure is confirmed by data collected by the Hospitality Association of Namibia (HAN), which recorded 588,086 guests in 2017. The 27 enterprises that participated in the business survey reported 26 beds on average. By applying this figure to the missing 20 values, we estimate that 456,980 overnight stays can potentially be sold per year. The average occupancy rate of the surveyed enterprises was 41.42 % (HAN data suggest an occupancy rate of 50.21 % for northern Namibia and 48.64 % nationwide). which means that 189,281 overnight stays were actually sold in 2017. The average duration of a stay in Zambezi hospitality enterprises was 3.15 days. Dividing the number of overnight stays by this factor yields an estimated 60,125 visitors in 2017, a figure that seems realistic when it is taken into consideration that both the number of establishments and the number of visitors have doubled since 2005. About one third of tourism in the region is domestic, while Europeans account for the largest group among the foreign visitors (table 2).

|  | Tourist arrivals in Namibia[1] | Tourism arrivals at north-eastern border posts[2] | Visitors in Zambezi accommodation facilities[3] | Visitors of Bwabwata National Park, eastern gate[4] |
|---|---|---|---|---|
| **Total no. of visitors** | 830,468 | 127,851 | 60,125 | 10,900 |
| **Namibian** | 29% |  | 34% | 17% |
| **RSA** | 12% | 12% | 6% | 13% |
| **Other African** | 2% | 40% | 20% | 2% |
| **European** | 48% | 30% | 24% | 56% |
| **US** | 4 % | 5% | 6% | 8% |

| **Other countries** | 5 % | 13 % | 10 % | 8 % |
|---|---|---|---|---|

*Table 2: Visitor counts in 2017. ¹HAN data, ²Own calculation, based on MET 2018 ³estimate, based on business survey, ⁴Suswe Gate, Bwabwata National park, from June 2018- June 2019*

Although the corridor serves primarily as a transport route from the resource extraction sites to Windhoek and the port at Walvis Bay, the development of infrastructure has brought about the intended promotion of tourism. During the high season, large shares of the traffic in the Zambezi region is tourism-related, which mainly benefits tour operators from Windhoek, but also tourism companies from elsewhere within the region. Yet, the Zambezi region serves merely as a stopover for tourists, as the low number of visitors in the national park indicates (table 2). The question arises as to whether the policy objectives of promoting growth in rural areas are being met, in other words, are rural households benefiting from tourism?

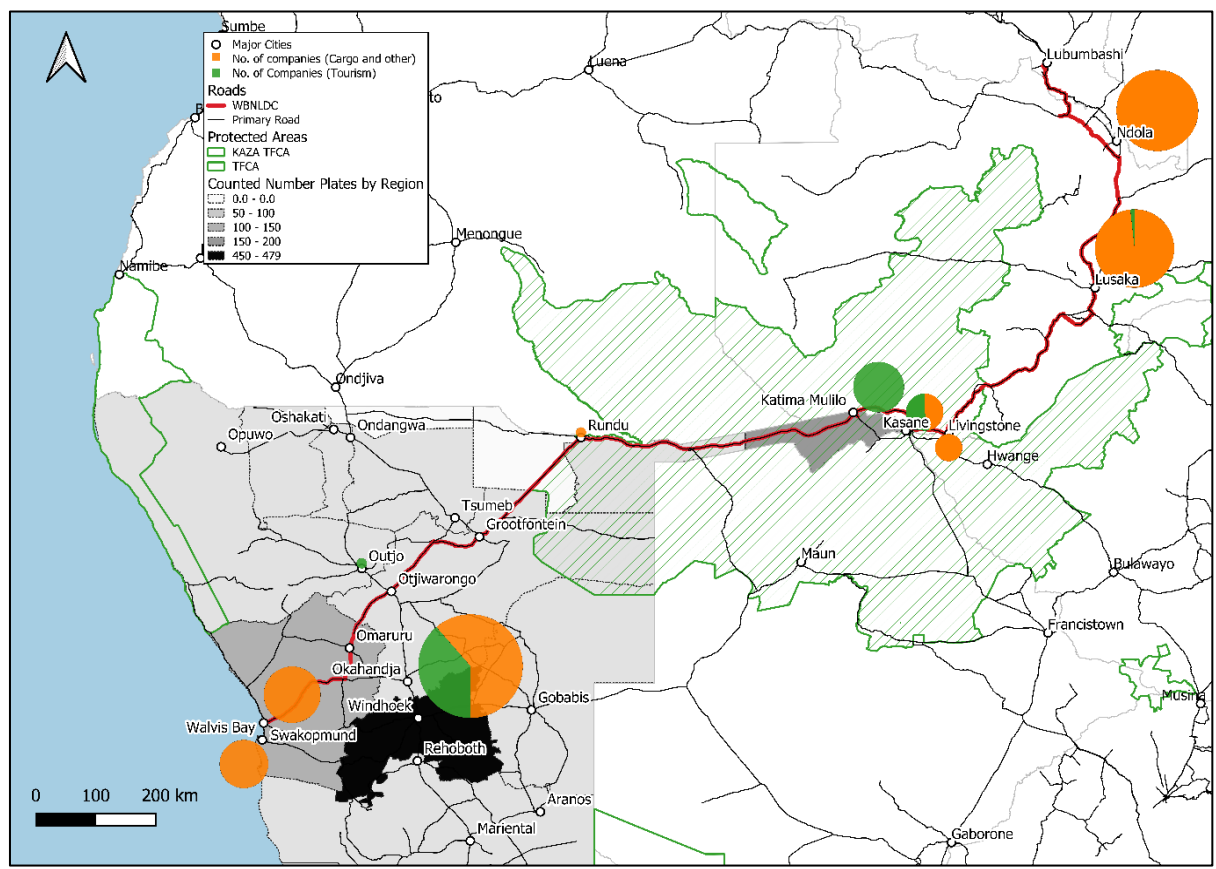

*Figure 4: Distribution of number plates and company headquarters along the corridor, based on traffic census data.*

### c. Who benefits? Local effects of a growth corridor (1000)

From a GPN perspective, regional development is closely bound to the ability of regions to capture value. A previous study had found that conservancies as local institutions are able to enforce value capture at local level, with roughly 20 % of the tourism value remaining in conservancies (Kalvelage et al., 2020). However, other literature suggests that conservancies are prone to elite capture (Silva and Mosimane, 2014) so value capture at conservancy level cannot be equated with an inclusive development strategy for large parts of the population. Therefore, to assess the effects of tourism on household income, household survey data are analysed.

As tourism is the primary source of funding for conservancies (over 95 %), tourism-related benefits for rural households in Zambezi can be classified as direct benefits (through employment at a lodge/ hunting camp or tourism business activity) and indirect benefits via conservancies (employment at a conservancy, cash pay-outs and other benefits, cf. table 2). There are additional indirect benefits that are not revealed by the data, for example the revenues generated from the supply of food and building material to lodges. However, a previous analysis found that the effect is small (Kalvelage et al., 2020).

*Table 3: Direct and indirect tourism-related benefits, Zambezi region. Household survey data*

| | Direct benefits | Indirect benefits via conservancies | | | | Total |
|---|---|---|---|---|---|---|
| **Benefit** | Employment at lodges | Employment at conservancy | Cash pay-out to households inside conservancies (last 12 months) | HWC offset-payments to households inside conservancies (last 12 months) | Non-monetary benefits for households inside conservancies | |
| **% of total workforce/ households (*No. of respondents*)** | 2.83 % (*41*) | 1.04 % (*15*) | 18.3 % (*64*) | 3.15 % (*11*) | 12.89 % (*45*) | |
| **Average amount (per month)** | 1,614 N$ | 1,196 N$ | 1,168 N$1 | 1,326 N$ | | |
| **Total amount (per year, whole rural Zambezi population)** | 15,105,023 N$ | 4,095,890 N$ | 1,310,864 N$ | 277,550 N$ | | 20,789,327 N$ (5.5 % of total household income) |

In the survey sample, it was not possible to identify any entrepreneur with a direct link to tourism. A previous study found that a lack of industry-specific skills, capital and a network constitute entry barriers for local entrepreneurs (Kalvelage et al., 2020). However, 41 respondents reported that they were employed by a tourism enterprise, which represents 2.83 % of the total workforce in the sample (1447, aged between 18 and 60). The average monthly wage is 1,614 N$ (114 USD). This low figure can be explained by the fact that the respondents are mainly employed in low-wage jobs, e.g. as waiters/waitresses, security guards or bartenders. Strikingly, the vast majority of the employees live in close vicinity to their place of work (97.5 %). According to the Zambezi 2011 census, the rural population figure for Zambezi is 62,234. Given that the sample represents 5.256 % of the population, the rural population of Zambezi earns roughly 15,105,023 N$ in annual wages from the tourism sector.

Projecting the number of jobs (41) to the whole population indicates that tourism enterprises in Zambezi provide 780 jobs, both formal and informal. Naidoo et al. (2016) found that lodges employ between 20 and 50 staff members and hunting camps 8-10. Given the total of 47 accommodation facilities and 11 hunting operators in the Zambezi region (Kalvelage et al., 2020), this would suggest an employment potential of between 1028 and 2460 employees in the tourism sector. However, the business survey includes smaller enterprises in urban areas, such as backpacker hostels with a considerably lower job creation effect. The finding based on the household survey therefore appears more accurate.

We add the indirect benefits resulting from the conservancy structure to these figures. The survey found that 15 individuals (or 1.04 % of the overall workforce) are employed by a conservancy, e.g. as game guards, enterprise officers or managers. The average monthly wage is 1,196 N$ (85 USD), which

adds up to a total contribution of household net income amounting to 4,095,890 N$ (250,000 USD) per year for the whole population. Furthermore, 75 households reported having received payments from a conservancy, either as part of a benefit-sharing programme (64) or as an Human-Wildlife Conflict (HWC) offset payment (11). These payments amount to 1,310,864 N$ (79,000 USD, cash pay-out) and 277,550 N$ (17,000 USD, HWC offset payments). 45 households reported having received non-monetary benefits, the most common being meat, electrification programmes and community funds (however, this is not included in the analysis). Totalling 20,789,327 N$ (1,471,850 USD), benefits derived from tourism for the Zambezi population as a whole represent 5.5 % of the net household income. In comparison, the Basic Social Grant is a monthly unconditional allowance of 1,100 N$ paid to all residents over the age of 60 ([https://www.social-protection.org/gimi/RessourcePDF.action?id=53959](https://www.social-protection.org/gimi/RessourcePDF.action?id=53959)). Our sample found n=388 household members aged 60 or older. Projecting this number to the whole population, we estimate that in the rural Zambezi region, 7382 elders receive a total of 81,202,000 N$ (5 m USD) per year.

Surprisingly, these figures are not in line with the results of an earlier report based on figures collected by Namibian Association of CBNRM Support Organisations (NACSO) (Kalvelage et al., 2020): According to NACSO data, conservancies in Zambezi paid a total of 400,000 USD in the form of direct cash pay-outs to members in 2017, while our findings suggest a total amount of 96,000 USD in 2019. Possible reasons for this discrepancy could be that the data were collected in different years or that there was a shift in policy from direct cash pay-outs to investments in development projects. However, as the figures differ quite strikingly, there is a need for further research. The data indicate that the intended benefits of conservation do not fully reach the conservancy members. In general, value capture from tourism at community level is low.

**Discussion and conclusion**

The brief historical background provided above makes clear that the economic opening of a region depends on infrastructure development. While the potential of the wildlife resource in the Zambezi region was already recognised in the early stages of colonial occupation, it required the creation of a conservation landscape and infrastructure development to couple the resource region to the tourism GPN. Parallel to infrastructure improvements, the steady expansion of the tourism sector in Zambezi can be observed. This process can be divided into three phases: first, the colonial era, when Zambezi possessed a peripheral status and was poorly connected to the rest of the country. During this time, wildlife was regarded as a potential resource, but its exploitation was limited to uncontrolled hunting activities. Second, under the apartheid regime, Zambezi was linked to the urban centres of South Africa and Namibia. The creation of a conservation imaginary laid the groundwork for exploiting the resource. Tourism in the region was still in its infancy, but expanded when conditions became more peaceful. Third, after independence, major efforts were made to improve the infrastructure connecting the region to Central Namibia, resulting in the tarring of the road and peaking in the construction of a bridge connecting Zambezi with Zambia. Simultaneously, new policies were developed to cement the region's status as a conservation territory, which led to far-reaching territorial claims.

Interview data reveal that the tarring of the road after independence was a major driver of tourism development in the region. However, the corridor development is a double-edged sword: while on the one hand, improved access to the region has facilitated tourism operations, nowadays Zambezi serves mainly as a stopover destination for travellers on their way to the main tourist attractions in the neighbouring countries due to reduced travelling time. Traffic census data suggest that most companies active in the region are based elsewhere. In the case of tourism, Windhoek is the gateway city, for other businesses, such as the logistics sector, the end nodes of the corridor (Walvis Bay and Swakopmund in the West, Lusaka and Ndola in the East) are the main beneficiaries of the corridor development.

This finding is supported by the analysis of household-level survey data: tourism-related revenues contribute little to the overall household income in the region. Less than 3 % of the respondents are employed in tourism, and entrepreneurial engagement with the tourism GPN is non-existent. Residents of conservancies additionally participate in the distribution of indirect benefits, though these are limited.

This paper aimed to examine the questions as to whether or not growth corridor policies fulfil the promise of fostering tourism in peripheral regions and, if this is indeed the case, whether local residents appropriate value. It has become clear that infrastructure development has kept its promise of increasing value creation from tourism in the region. However, the corridor serves mainly to connect tourism companies in the gateway to the resource region. By envisioning tourism-driven development, the GPN fosters territorialisation in the resource region, thereby ensuring the maintenance of a conservation landscape with the result that actors in Zambezi are suppliers to lead firms in the global nodes of the tourism GPN, including Windhoek. By means of infrastructure development, Zambezi is integrated into the GPN, but serves merely as a resource region. For this reason, the promises of rural development reach only a very limited number of people who are employed in low-wage jobs and/ or receive payments from conservancy managements. However, it is necessary to point out that this paper examines the intended development effects of conservancy and growth corridor policies. While these effects may be less than what was aspired to, this paper does not address conservation successes. Wildlife tourism is an industry that is based on conservation and thus has an interest in preserving the ecosystem. Despite the limited direct household benefits, the tourism industry expands the national tax base, which in return, benefits households in Zambezi through social transfers, e.g. the Basic Social Grant. Alternative development paths based on the extractive sector, such as the fracking of oil, which has recently been the topic of debate, have detrimental effects on the ecosystem and research suggests that local benefits are very limited. Therefore, local residents' participation in the tourism industry needs to be strengthened and alternative income opportunities that can be combined with the overall aim of nature conservation need to be explored.

Integrating these findings into a broader debate on GPNs, it can be postulated, first, that territorialisation is a key for value creation in nature-based GPNs. As the input to the GPN depends on "biophysical processes prior to human labour" (Bridge, 2008), exerting control over the area where the resources are located is a precondition for value creation. In the case of tourism in Zambezi, control is exerted by creating a territorial conservation imaginary, which is achieved in the creation of nature reserves and conservancy legislation; and which occupies large parts of the area. Second, the resource that lays the groundwork for the tourism GPN is wildlife and wilderness landscapes and this resource is place-bound and not easily transferable. It is therefore crucial to provide infrastructure access in order to couple with the GPN and thus to achieve value creation.

Furthermore, the case presented above illustrates the territoriality of GPNs once again. Coupling with a GPN does not necessarily have positive outcomes in the resource region, but development effects emerge in urban agglomerations along the growth corridor. These findings have two major implications for policy-making: first, when promoting a sector in a region, it is necessary to consider the territorial claim that accompanies it, which may potentially displace alternative development pathways. Second, growth corridor policies need to ensure that the envisioned growth materialises not only in the urban centres, but also involves actors in the resource region. The integration of resource regions into GPNs should include policies that foster active entrepreneurial engagement with the GPN, thus ensuring the capture of value on a regional scale to prevent the formation of enclave-like economic structures.